\documentclass[final,5p,times,twocolumn]{elsarticle}

\usepackage{graphicx}
\usepackage{amssymb}
\usepackage{lineno}
\usepackage[euler]{textgreek}
\usepackage[amssymb,thinqspace,thinspace]{SIunits}
\usepackage{color}
\journal{Materials Science and Engineering A}

\begin{document}
\modulolinenumbers[5]
\begin{frontmatter}


\title{Fatigue cracking in gamma titanium aluminide}
%Effect of temperature excursions on fatigue crack growth in...



\author[IC]{Claire F. Trant}
\author[IC]{Trevor C. Lindley}
\author[RR]{Nigel Martin}
\author[RR]{Mark Dixon}
\author[IC]{David Dye}

\address[IC]{Department of Materials, Royal School of Mines, Imperial College, Prince Consort Road, London, SW7 2BP, UK}
\address[RR]{Rolls-Royce plc., PO Box 31, Derby, DE24 8BJ, UK}

\begin{abstract}
\textcolor{red}{Cast and HIP'ed \textgamma-TiAl 4522XD is being developed for use in jet engine low pressure turbine blades, where temperature variations occur through the flight cycle.  The effects of temperature variations on fatigue cracking were therefore examined in this study.}
It was found that fatigue crack growth rates were higher at $750\celsius$ than $400\celsius$, but that $\Delta K_\mathrm{th}$ was also higher. Temperature excursions between $400$ and $750\celsius$ during fatigue crack growth resulted in retardation of the crack growth rate, both on heating and cooling.
It was also found that for notches $0.6$~mm in length and smaller, initiation from the microstructure could instead be observed at stresses similar to the material failure stress; a microstructural initiation site exists. 
A change from trans- to mixed trans-, inter- and intra-lamellar cracking could be observed where the estimated size of the crack tip plastic zone exceeded the colony size.
\end{abstract}
\end{frontmatter}
%\linenumbers

\section{Introduction}
$\gamma$-TiAl based alloys have long been attractive for use in jet engines due to their low density, around $4\usk\gram\usk\centi\meter^{-3}$, strength, oxidation resistance and creep resistance at intermediate temperatures in the range of $450$--$700\celsius$~\citep{Clemens2013}. 
In common with most intermetallics, this is because their ductility and toughness are quite limited, due to their large unit cells and hence dislocation Burgers vector lengths. This has prompted development efforts stretching back as far as 1953 \cite{Williamsa}, but only recently have this class of alloys begun to see service in civil jet engines \cite{FritzAppel2011}. Commercial alloys in this class largely use a two-phase structure consisting of a majority tetragonal L$1_0$ $\gamma$-TiAl phase with some hexagonal DO$_{19}$ $\alpha_2$-Ti$_3$Al phase. In cast microstructures the TiB$_2$ phase is used as a grain refiner of the prior-$\beta$ grains \cite{LARSEN199145}, ($\beta$ is the \emph{bcc} high temperature solid solution phase of Ti), with a fully lamellar $\gamma+\alpha_2$ microstructure obtained within these grains. Whilst these microstructures are not ductile at room temperature, they are found to be relatively tough, which enables their use in gas turbine components where some resistance to fatigue crack initiation is required, \emph{e.g.} around features, foreign object damage or manufacturing imperfections. 
The two phase multilayer lamellar system within the grains shows improved mechanical properties when the lamellae are of a reduced thickness \cite{kruzic1999gamma,MINE201213}. Due to the tendency for interfaces to take a preference for the lowest interfacial energy, there is a common  orientation relationship between the phases at the lamellar boundaries. The matching close packed planes and directions are
$\{111\}_\gamma \parallel(0001)_{\alpha_2}$ \hspace{0.1mm} and $<$1$\bar{1}$0]$_{\gamma}$ $\parallel$ $<$11$\bar{2}0>_{\alpha_2}$
\cite{FritzAppel2011}. Due to the symmetry of the \textgamma\hspace{0.1mm} phase there are six orientation variants. Due to these six rotation variants of the \textgamma\hspace{0.1mm} phase, the \textgamma~$\parallel$~\textgamma\hspace{0,1mm} boundaries can be pseudo twins with a 60\degree\hspace{0.1mm} misorientation, 120\degree\hspace{0.1mm} rotational faults or true twins with 180\degree\hspace{0.1mm} rotation, from the highest to lowest interfacial energy variant respectively \cite{APPEL1998187}.

Overall ductility in both the $\alpha_2$ phase and the $\gamma$ phase independently increase with increasing temperature from room temperature (RT) up to $800\celsius$ \cite{kim1994ordered}, as well as the ductility of the overall alloy. Deformation at RT is mainly confined to the \textgamma\hspace{0.1mm} phase of the two phase alloy \cite{singh2006activation,appel1993deformation,appel1998microstructure,vasudevan1989influence,sriram1997geometry,wiezorek1998deformation,singh2006situ}, deforming by octahedral glide of ordinary dislocations with the Burgers vector \textbf{b}=1/2$<$110] \cite{Shechtman1974} and super dislocations with the Burgers vectors \textbf{b}=$<$101] and \textbf{b}=1/2 $<$11$\bar{2}$] \cite{Lipsitt1975,doi:10.1080/01418618608242882}. Mechanical twinning in the \textgamma \hspace{0.1mm} is a secondary deformation mode \cite{singh2006situ}. $\alpha_{2}$ is seen to deform less, attributed to its increased solubility for alloying elements compared to the \textgamma \hspace{0.1mm} phase. This is understood to happen because dislocations are more easily pinned, \emph{e.g.} at solute atoms, and so its plasticity is more limited \cite{FritzAppel2011}.

Deformation at $800\celsius$ is mainly transferred by the $\gamma$ phase due to glide modes similar to those occurring at RT, with the addition of increased strain accommodation from climb of ordinary dislocations \cite{kad1994contribution,yoo2002nonbasal}, relaxing the need for independent slip systems. $\gamma$ also has increased mechanical twinning at higher temperatures. Increasing temperature increases the activation of prismatic, basal and pyramidal slip in \textalpha$_2$, with ease of activation in that order \cite{UMAKOSHI19931149}. Pyramidal slip can be seen to decrease at intermediate temperatures \cite{FritzAppel2011}. Due to pyramidal slip being the only glide plane to have a \textit{c} component, it is required for general plasticity of the material. However, due to prismatic slip increasing at higher temperatures, and pyramidal slip decreasing at intermediate temperatures, this causes a strong preference for prismatic slip, and can contribute to a decrease in the ductility of the \textalpha$_2$ phase at increased temperatures \cite{inui1993plastic,minonishi1995dissociation,minonishi1993intermetallic,umakoshi1993high}. The increase in number of available deformation mechanisms in both phases reduces the occurrence of interfacial incompatibility between deformation in each of the the two phases and contributes to increased ductility at increased temperatures, due to improved slip transfer.

\textcolor{red}{Fracture behaviour of the \textgamma \hspace{0.1mm} and \textalpha$_2$ phases is temperature dependent, and therefore fatigue crack growth characteristics can be seen to altered with a change in temperature. Crack paths are seen to be less tortuous at elevated temperatures. This is believed to be due to activation of a higher number of slip systems. Change in temperature results in homogenisation of slip, which is assisted by diffusion and cross glide. The change in deformation modes with temperature causes variations in crack growth rates in comparison to those seen at room temperature. At elevated temperature the tendency for planar slip is no longer observed, in contrast to room temperature \cite{Aswath1991}. Overall, dominant fracture mechanisms present in lamellar alloys are interfacial delamination, trans-lamellar fracture and decohesion of lamellar colonies \cite{Leyens2003}.}

\textcolor{red}{The lath width of lamellar microstructures has an effect on fatigue threshold and crack propagation \cite{MINE}; microstructures with smaller lath spacings possess a higher threshold than the larger spacings. This increase in resistance to fatigue crack growth is attributed to the number of lamella interfaces per unit distance on the crack path. Crack growth rates in \textgamma-TiAl are increased by the presence of a sharp pre-crack \cite{Suresh1991}. Conversely, crack tip oxidation above 700\degree C \cite{mercer}, can result in oxide-induced crack closure.}

It remains the case, however, that the limited ductility and relatively fast crack propagation of $\gamma$-TiAl make its use as a material for highly stressed components particularly challenging. Fatigue crack growth threshold $\Delta K_\mathrm{th}$ has been identified as a key material property for design, supported by an improved understanding of the deformation mechanisms associated with a crack growing near-threshold. For example, cracks can be seen to be locally arrested by a change in colony orientation~\cite{Dahar2015}.  Excursions in temperature may also be of interest, as the $\gamma$ phase can activate additional deformation mechanisms at around 750\degree C, rendering it ductile.  Therefore in this paper, the effect of temperature excursions on the fatigue crack growth behaviour and associated fracture surface features are examined in a commercial $\gamma$-TiAl alloy, cast 4522XD  (Ti-45Al-2Mn-2Nb-0.8 vol.\% TiB$_2$-XD). 

\section{Experimental Procedures}

\begin{figure}[t!]
\centering\includegraphics[width=1\linewidth]{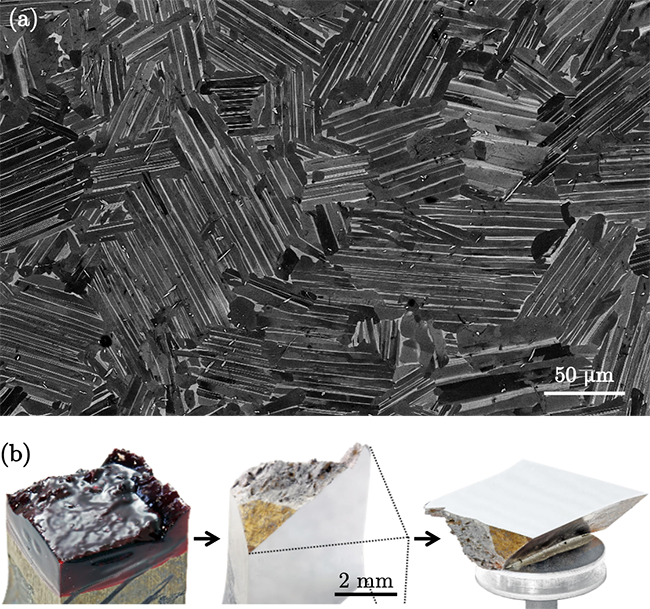}
\caption{(a) Microstructure of as-cast, HIP'ed \textgamma-TiAl. (b) Sectioning, polishing and mounting of the fractured specimen for SEM and EBSD. Images courtesy of Tom Zalewski, Rolls-Royce plc..} \label{fig:micro}
\end{figure}

The alloy composition is provided in Table~\ref{tab:compo}. The as-cast and HIP'ed microstructure is shown in Figure~\ref{fig:micro}(a).  Within the prior-$\beta$ grains randomly oriented $\gamma+\alpha_2$ colonies were found, 40-$100\usk\micro\meter$ in size, consisting of 92\% fraction of $\gamma$ lamellae, 0.5~-~3.6 \textmu m in plate width, averaging 1.4 \textmu m, often with several plates together. The \textgamma \hspace{0.1mm} is interleaved with thin ligaments of \textalpha$_2$, 0.2 - 0.5 \textmu m in plate width, averaging 0.3 \textmu m \cite{Edwards2016a}.

\begin{table}[hbt!]
\centering
\begin{small}\begin{tabular}{ccccccc}
Ti & Al&Mn&Nb&B&Si&O\\
bal. &43.90&1.80&1.90&0.90&0.20&0.14 \\
\end{tabular}\end{small}
\caption{Composition of the 4522XD alloy studied, in at.\%, measured by ICP-OES and LECO analysis. 2 vol.\% TiB$_2$ was added to the melt, and the H content was found to be $<10\usk$ppmw.}\label{tab:compo}
\end{table}

\begin{figure}[b!]
\centering
\includegraphics[width=1\linewidth]{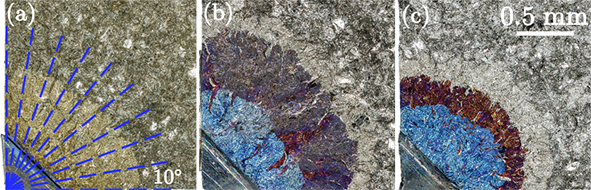}
\begin{small}\begin{tabular}{l |l|l }
\hline
\textbf{Test}  &\textbf{`Beach marks'} & \textbf{Held at} \\
\hline
a \hspace{0.1cm}1 & 109 \textmu V growth, 400\degree C for 260 mins  & Mean load 
\\
\hline
b \hspace{0.12cm}1 & 60 \textmu V growth, 650\degree C for 240 mins & Max load
\\
\hspace{0.3cm}2 & 170 \textmu V growth, 550\degree C for 240 mins  & Max load
\\
\hspace{0.3cm}3 & 70 \textmu V growth, overload & Max load
\\
\hline
c \hspace{0.1cm}1 & 78 \textmu V growth, 650\degree C for 240 mins   & Max load
\\
 \hspace{0.3cm}2 & 79 \textmu V growth, 550\degree C for 240 mins  & Max load
 \\
 \hspace{0.3cm}3 & 80 \textmu V growth, overload & Max load
\\

\hline
\end{tabular}\end{small}
\caption{The crack front was measured every 10\degree \hspace{0.1cm} and averaged, overlaid on cropped fracture surface of (a) calibration a with only one data point. (b) Calibration b with three crack fronts, (c) calibration c with three crack fronts. Photography courtesy of Failure Investigation, Derby, Rolls-Royce plc.. Bottom, table of methods used to generate the features on the fracture surfaces.}\label{Figure:Calibration}
\end{figure}

\begin{figure*}[t!]
\centering\includegraphics[width=0.89\textwidth]{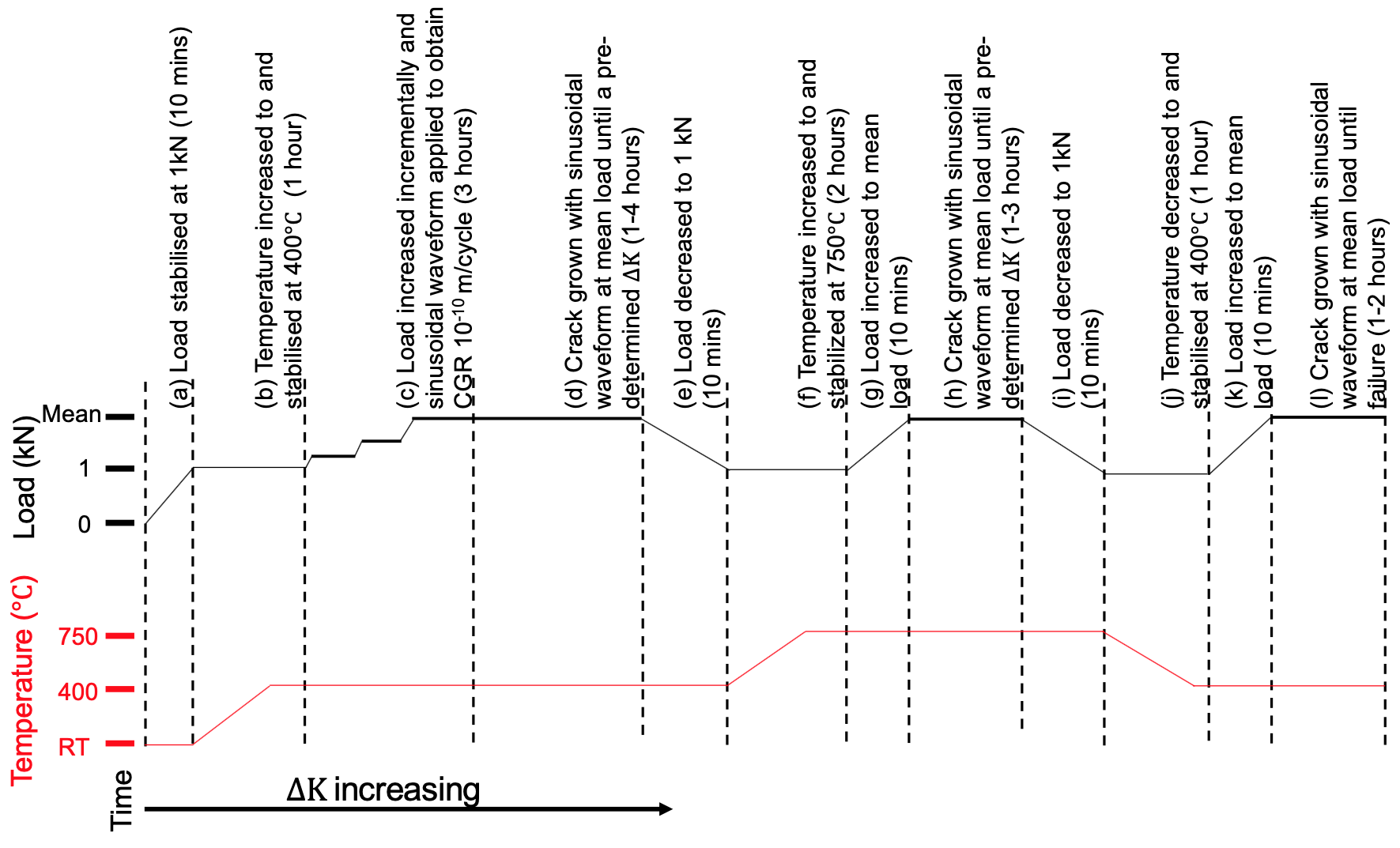}
\caption{\textcolor{red}{Evolution of temperature and maximum load during the fatigue testing through the test sequence, showing how the changes in temperature were applied, with approximate time taken for each step.}} \label{figure:test}
\end{figure*}

\begin{figure}[b!]
\centering\includegraphics[width=1\linewidth]{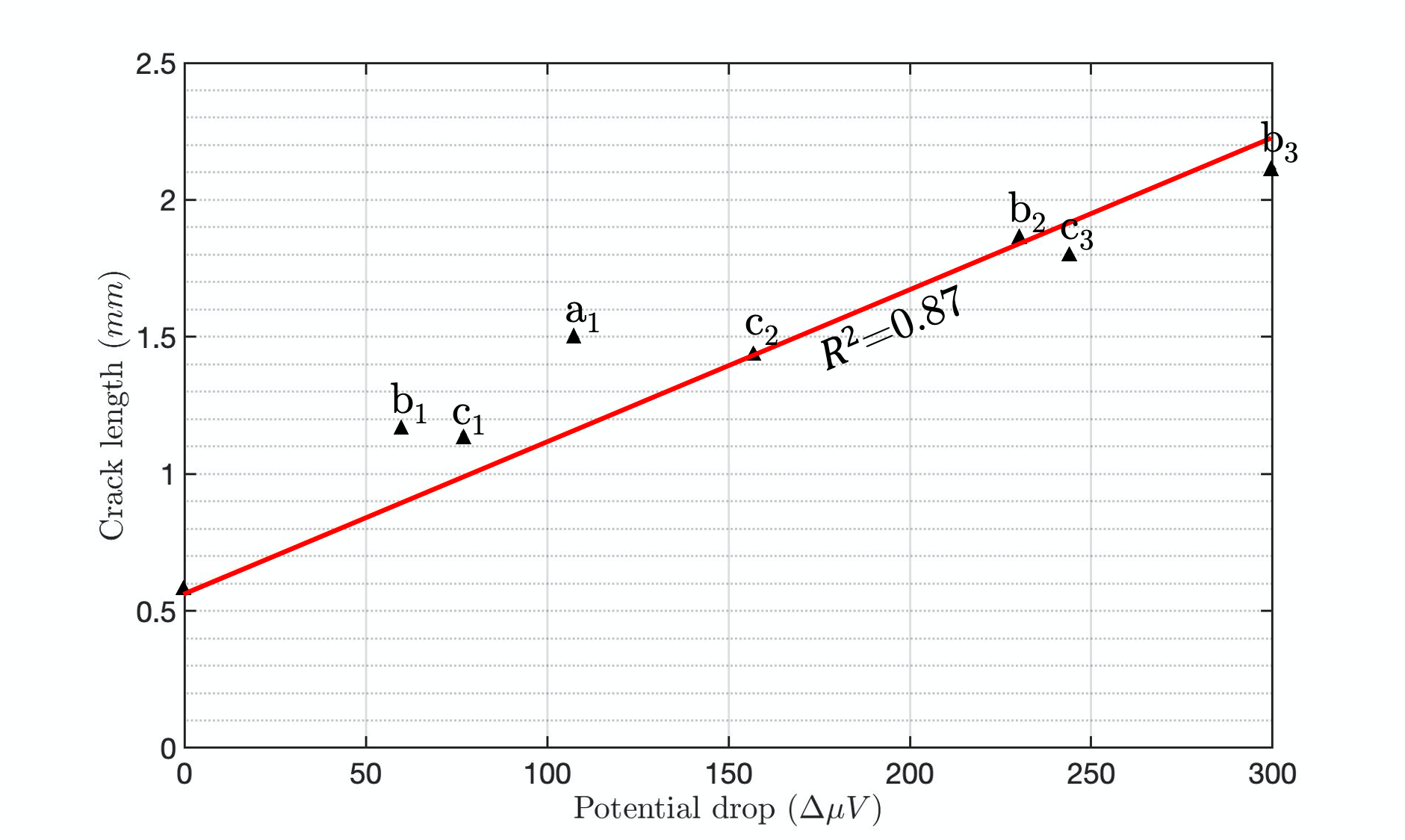}
\caption{Data points from calibration samples and the zero datum point, to give the DCPD calibration used.} \label{fig:curve}
\end{figure}

Cast bars were machined to form M12 threaded $5\times 5\usk\milli\meter$ square section specimens with gauge length $15\usk\milli\meter$  with a machined corner notch, $0.5\usk\milli\meter$ in depth\textcolor{red}{, (measured at $45\degree$ to the specimen edges, with the notch shown in Figure 2a), \cite{Yang,iso}. 0.5, 0.25, 0.1 and 0.06 mm notches were machined with a diamond saw. The shape of the as-machined notches can be seen to form a U shape, with a crack notch radius of $\sim$50 $\micro\meter$. This was considered to be a sharp notch \cite{christman1986crack,Pippan1994}. In addition, a 0.03~mm notch was machined edge-on using a gallium Focussed Ion Beam (FIB) on a FEI Helios Nanolab 600 DualBeam system.}

The specimen gauge lengths were then hand polished using OPS. Indirect crack growth measurement was carried out using the direct current potential drop (DCPD) technique. The loading waveform was sinusoidal, R=0.1, at a frequency of 5~Hz. Measuring fatigue crack growth of small cracks required 2 x 0.063 mm diameter probes spot welded either side of the specimen notch. The specimen was inserted and the load stabilised at 1 kN, the temperature was then increased and stabilised. Threshold was determined using incremental increases in $\Delta K $ until a crack growth rate of 10\textsuperscript{-10}~m / cycle was observed over 10\textsuperscript{4} cycles. For tests with a varied temperature, the crack growth was paused at a pre-determined $\Delta K$.  The loading waveform was paused, load decreased to a nominal value and the temperature increased, stabilised within $\pm 0.5\%$ of  the desired temperature, using a three-zone radiant furnace. The temperature was measured using a thermocouple located on the shoulder of the sample. The load was then returned to the mean value, and the loading waveform re-applied. The crack was then left to grow until a larger, pre-determined $\Delta K$. The waveform was again paused, load decreased, and temperature returned to the original temperature, and left to stabilise. The same waveform was again re-applied and the crack allowed to grow until failure, Figure 3. Resulting fracture surfaces were then polished perpendicular through the face diagonal, again using OPS, in preparation for SEM and EBSD, Figure~\ref{fig:micro}(b).

%\section{Calibration}between the crack le
The relationship between the crack length and the potential drop must be determined via calibration. This requires creating features in the fracture surface at a known potential drop. \textgamma-(TiAl) does not form striations when undergoing an overload, so calibration was carried out using heat tinting of the fracture surface. This was carried out at various time intervals in order to allow oxidation, leaving `beach marks' of decreasing oxidation thickness, summarised in Figure \ref{Figure:Calibration}. The specimen was then held at load so as to hold open the crack to allow oxidation. The crack length of each `beach mark' was measured every 10\degree \hspace{0.1mm} and an average taken of these values. Figure \ref{Figure:Calibration} gives seven `beach marks' on three specimens, therefore seven points for the calibration curve. The datum point for zero potential drop can also be included, with the crack length of the starter notch.  These give the calibration curve, Figure \ref{fig:curve}, with the data points labelled with their relative `beach mark'. The calibration curve was created in accordance with \cite{RR}. The calculations for the change in stress intensity factor  $\Delta K$ for the corner crack specimen used throughout can be found in \cite{iso, BSI2010}.
% \begin{table}[b!]
% \centering
% \begin{small}\begin{tabular}{l |l | l}
% \hline
% \textbf{Test no.}  &\textbf{Method} & \textbf{Held at} \\
% \hline
%  1 & 29 \textmu V growth, 400\degree C for 3 hours & Mean load 
%  \\
%   &  40 \textmu V growth, 400\degree C for 1hr 20min 
%  \\
%   &  40 \textmu V growth, 400\degree C for 1hr
% \\
% \hline
% 2  & 60 \textmu V growth, 650\degree C for 4 hours & Max load
% \\
% & 170 \textmu V growth, 550\degree C for 4 hours 
% \\
% & 70 \textmu V growth, overload
% \\
% \hline
% 3 & 78 \textmu V growth, 650\degree C for 4 hours  & Max load
% \\
%   &79 \textmu V growth, 550\degree C for 4 hours 
%  \\
%   & 80 \textmu V growth, overload 
% \\

% \hline
% \end{tabular}\end{small}
% \caption{The methods used to generate features on the fracture surface to create a calibration curve.}\label{label:Calibration}
% \end{table}
% \begin{figure}[b!]
% \centering\includegraphics[width=1\linewidth]{Calibration.png}
% \caption{Measurements of crack front taken every 10\degree \hspace{0.1} and averaged, overlaid on cropped fracture surfaces of (a) calibration a with only one data point (b) calibration b with three crack fronts (c) calibration c with three crack fronts. Photography courtesy of Failure Investigation, Derby, Rolls-Royce plc..} \label{fig:calib}
% \end{figure}

\section{Results}
\subsection{Temperature Excursions}
\begin{figure}[b!]
\centering\includegraphics[width=1\linewidth]{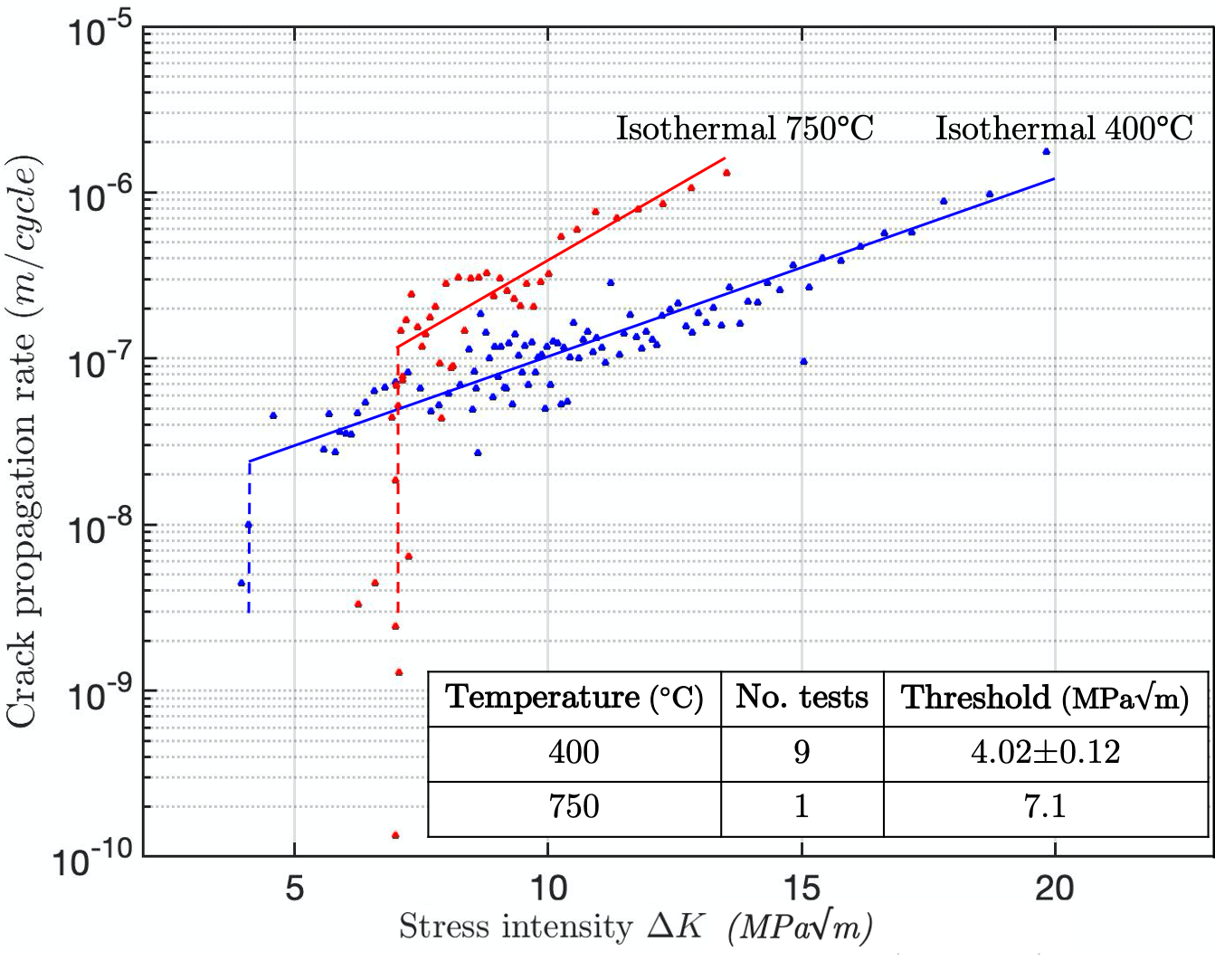}
\caption{\textcolor{red}{Two isothermal fatigue crack growth curves obtained at (blue dots) 400 and (red dots) 750\degree C in \textgamma-(TiAl) alloy 4522XD in the as-cast and HIP condition.  The lines are guides to the eye, added to aid the reader. The test at 400\degree C a representative instance of the nine tests performed.}} \label{fig:isothermal}
\end{figure}

Isothermal response fatigue crack growth curves at $400$ and $750 \celsius$ are shown in Figure~\ref{fig:isothermal}, with a guide to the eye in a solid line for the Paris region and a dotted line for the threshold. At 400\degree C $\Delta K_\mathrm{th}$ in \textcolor{red}{a predominantly} trans-lamellar fracture mode was $4.02\pm 0.12\quad \mathrm{MPa}\sqrt{\mathrm{m}}$ from nine specimens. \textcolor{red}{In subsequent sections, variations in notch depth are also examined, with similar results. This level of repeatability in threshold determination is regarded as quite high.}  At an increased temperature of $750\celsius$, trans-lamellar $\Delta K_\mathrm{th}$ increases to 7.1~MPa$\sqrt{\mathrm{m}}$. \textcolor{red}{Although this is from just one test, given the repeatability at $400\celsius$, this is regarded as good evidence that threshold is higher at $750\celsius$ than at $400\celsius$.} This was tested using 0.5 mm notched specimens, and therefore initiating through multiple randomly oriented colonies local to the notch and therefore in trans-lamellar fracture mode. The fatigue crack growth rates in the steady-state (Paris) fatigue crack growth regime are observed to be greater at 750\degree C than 400\degree C, with a concomitant reduction in life. 

Four tests were carried out in which the temperature was increased from 400 to 750\degree C for differing ranges of $\Delta K$, summarised in Figure 6. Each of the fatigue crack growth curves show blue data points at 400\degree C and red data points at 750\degree C. This data is overlain with a blue line for the isothermal response curve at 400\degree C and a red line for the isothermal response curve at 750\degree C, from Figure 5.

% \begin{table}[h!]
% \centering
% \begin{small}\begin{tabular}{l |l | l }
% \hline
% \textbf{Test} & \textbf{$\Delta$K temperature increase } &\textbf{$\Delta$K temperature decrease  }  \\
% \textbf{} &  \hspace{14mm}(MPa$\sqrt{\mathrm{m}}$) & \hspace{14mm}(MPa$\sqrt{\mathrm{m}}$) \\
% \hline
%  \hspace{3mm}A & \hspace{21mm}7.8 & \hspace{21mm}9.4
%  \\
%  \hspace{3mm}B &\hspace{21mm}7.0 & \hspace{21mm}11.1 
%  \\
%  \hspace{3mm}C & \hspace{21mm}5.6 & \hspace{21mm}8.5 
% \\
%  \hspace{3mm}D & \hspace{21mm}7.4 & \hspace{21mm}n/a \\
 
% \hline
% \end{tabular}\end{small}
% \caption{Table of tests and their values of $\Delta K$ for temperature increase to 750\degree C and decrease to 400\degree C.}\label{table:param}
% \end{table}

% \begin{figure}[h!]
% \centering\includegraphics[width=.92\linewidth]{TE1.png}
% \centering\includegraphics[width=.92\linewidth]{TE3.png}
% \centering\includegraphics[width=.92\linewidth]{TE2.png}
% \centering\includegraphics[width=.92\linewidth]{TE4.png}
% \caption{Fatigue crack growth test 1, with a temperature increase from 400 to 750 \degree C for the interval from 7.8 - 9.4 MPa$\sqrt{\mathrm{m}}$. The isothermal behaviour guides-to-the-eye are overlain.} \label{fig:TE1}
% \end{figure}

\begin{figure}[h!]
\centering\includegraphics[width=0.77\linewidth]{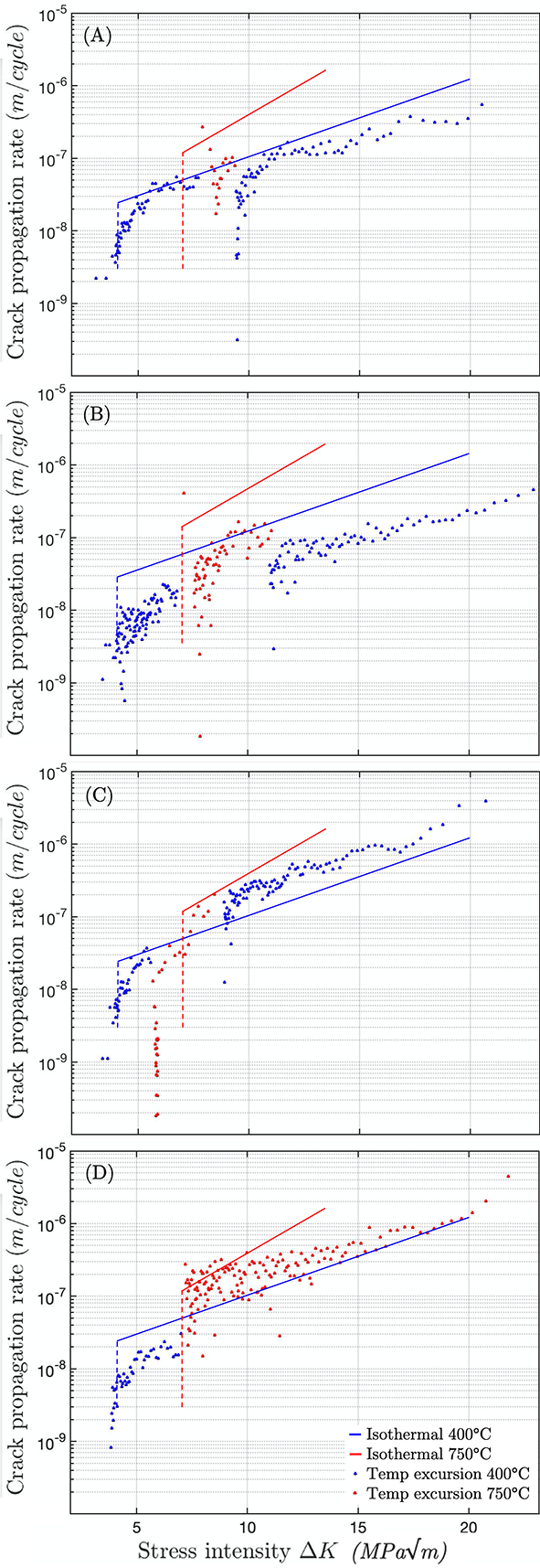}

\vspace{2mm}

\begin{footnotesize}\begin{tabular}{l |l | l }
\hline
\textbf{Test} & \textbf{$\Delta$K temperature increase } &\textbf{$\Delta$K temperature decrease  }  \\
\textbf{} &  \hspace{10mm}(MPa$\sqrt{\mathrm{m}}$) & \hspace{10mm}(MPa$\sqrt{\mathrm{m}}$) \\
\hline
 \hspace{1mm}A & \hspace{15mm}7.8 & \hspace{14mm}9.4
 \\
 \hspace{1mm}B &\hspace{15mm}7.0 & \hspace{13mm}11.1 
 \\
 \hspace{1mm}C & \hspace{15mm}5.6 & \hspace{14mm}8.5 
\\
 \hspace{1mm}D & \hspace{15mm}7.4 & \hspace{14mm}n/a \\
 
\hline
\end{tabular}\end{footnotesize}

\caption{Fatigue crack growth tests, with a temperature increase from 400 to 750\degree C. The isothermal behaviour guides-to-the-eye are overlain. } \label{fig:tes}
\end{figure}

% (a) Test A, with a temperature increase for the interval from 7.8 - 9.4 MPa$\sqrt{\mathrm{m}}$. (b) Test C, with a temperature increase for the interval from 7.0 - 11.1 MPa$\sqrt{\mathrm{m}}$. (c) Test C, with a temperature increase for the interval from 5.6 - 8.5 MPa$\sqrt{\mathrm{m}}$. (d) Test D, with a temperature increase after 7.4 MPa$\sqrt{\mathrm{m}}$.

\textcolor{red}{In test A, Figure \ref{fig:tes}(A), the temperature was increased from $400\celsius$ to $750\celsius$ for crack growth between 7.8 MPa$\sqrt{\mathrm{m}}$ and 9.4~MPa$\sqrt{\mathrm{m}}$.} In the increased temperature region there is seen to be an initial burst in growth, showing a high crack growth rate. %This is thought to be due to the crack growing suddenly, in an inter-lamellar fracture mode through a colony. This high crack growth rate is short lived. The fracture path is then impeded by the following colony. 
The $\Delta K$ is then close to the trans-lamellar fatigue crack growth threshold at 750\degree C in air, and therefore the crack growth rate \textcolor{red}{reduces, e.g. due to crack tip blunting, and} then stabilises. It is uncertain whether it would eventually grow out to equilibrium for 750\degree C. When the temperature is then decreased again, the crack growth rate retards due to a larger plastic zone size and larger yield stress, 430~MPa at 400\degree C and 350 MPa at 750\degree C. This slowly grows out to equilibrium to the isothermal response curve for 400\degree C. 

Test B, Figure \ref{fig:tes}(B), can be seen to give similar results, with the elevated temperature portion in the $\Delta K$ range from 7.0~MPa$\sqrt{\mathrm{m}}$ to 11.1 MPa$\sqrt{\mathrm{m}}$. In the increased temperature region, this test can again be seen to show an initial burst of growth.% due to growing in inter-lamellar fracture mode, and then being impeded by the following colony. 
The crack growth rate grows out to equilibrium more slowly than test A due to the stress intensity factor range being closer to the trans-lamellar fracture mode threshold at 750\degree C in air. Test B again gives no indication that the crack growth rate would have grown out to the isothermal response curve at 750\degree C.

%\begin{figure}[h]
%\centering\includegraphics[width=1\linewidth]{TE3.png}
%\caption{Fatigue crack growth test 3, with a temperature increase from 400 to 750 \degree C for the interval from 7.0 - 11.1 MPa$\sqrt{\mathrm{m}}$. The isothermal behaviour guides-to-the-eye are overlain.} \label{fig:TE3}
%\end{figure}

In test C, Figure \ref{fig:tes}(C), the increase to $750\celsius$ was performed below the isothermal fatigue crack growth threshold in trans-lamellar fracture mode at $750\celsius$. The temperature was increased from 400 to 750\degree C at $\Delta K$ of 5.6 MPa$\sqrt{\mathrm{m}}$, but trans-lamellar $\Delta K_\mathrm{th}$ is 7.1 MPa$\sqrt{\mathrm{m}}$ at 750\degree C. This caused the crack growth to retard by a multiple of 10\textsuperscript{3}. However, due to the sharp crack already initiated, the crack slowly grows out. This gives the fatigue crack growth threshold \textcolor{red}{at 750\degree C in air given a sharp pre-crack to be lower, at 5.8~MPa$\sqrt{\mathrm{m}}$. It is again uncertain whether this elevated temperature region would have grown out to the isothermal} $750\celsius$ fatigue crack growth curve. Once the temperature was decreased again at $\Delta K$ of 12~MPa$\sqrt{\mathrm{m}}$ the bigger plastic zone size again retards crack growth. 

In test D, Figure \ref{fig:tes}(D), the temperature was increased to 750\degree C at a $\Delta K$  of 7.4 MPa$\sqrt{\mathrm{m}}$ until failure, without returning to 400\degree C. There is large variability in the crack propagation rate for the same material in the same test environment. In Figures~\ref{fig:tes}(A)--(D) it can be seen that when the temperature is increased from 400 to 750\degree C, whilst the isothermal response of Figure 5 is not recovered, the grown-out crack growth rate is always higher than at 400 \degree C, albeit with some variability due to the relatively coarse prior-$\beta$ grain size. %Therefore whilst complete recovery to the isothermal case does not occur in the length of crack available in the short corner crack samples used, it is suggested that it is likely that these eventually would\footnote{The choice of specimen geometry was a consequence of matching the casting conditions closely to those that might be expected to be used in aerofoils, i.e. limited sample section size.}.

\begin{figure}[t!]
\centering\includegraphics[width=1\linewidth]{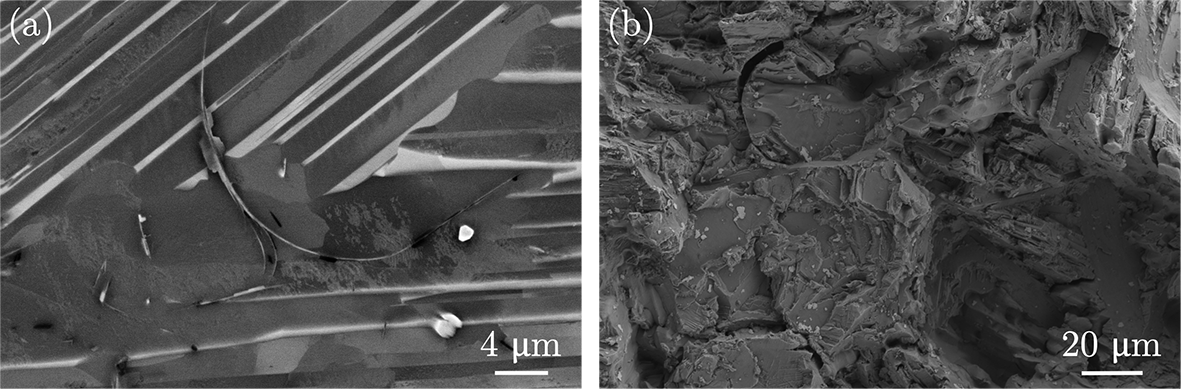}
\caption{(a) Backscattered electron image of a wavy boride. (b) Secondary cracking on the fracture surface around a boride.} \label{fig:2crack}
\end{figure}

%\begin{figure}[h]
%\centering\includegraphics[width=1\linewidth]{TE2.png}
%\caption{Fatigue crack growth test 2, with a temperature increase from 400 to 750 \degree C for the interval from 5.6 - 8.5 MPa$\sqrt{\mathrm{m}}$. The isothermal behaviour guides-to-the-eye are overlain.} \label{fig:TE2}
%\end{figure}
%\begin{figure}[h]
%\centering\includegraphics[width=1\linewidth]{TE4.png}
%\caption{Fatigue crack growth test 4, with a temperature increase from 400 to 750 \degree C after 7.4 MPa$\sqrt{\mathrm{m}}$. The isothermal behaviour guides-to-the-eye are overlain.} \label{fig:TE4}
%\end{figure}
Around a crack tip bifurcation of the crack may often take place, but as these link up to provide overall crack advance this may leave behind cracked ligaments that did not ultimately propagate. Furthermore, residual stresses and closure effects may also lead to cracking in the plastic wake. This can occur during any of the cracking modes, around grain boundaries, or through borides. The grain boundary borides were wavy in appearance, as shown in Figure \ref{fig:2crack}(a), and are associated with such secondary cracking, as can be seen in Figure \ref{fig:2crack}(b).  \textcolor{red}{Such secondary cracking could be observed across the fracture surface; one might speculate that this occurred more in the regions of retardation associated with temperature changes.}

\begin{figure}[b!]
\centering\includegraphics[width=1\linewidth]{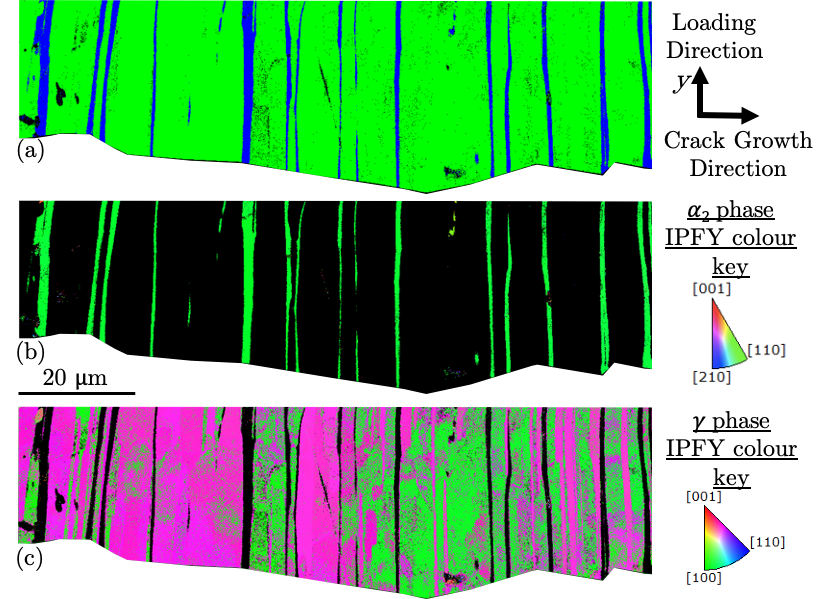}
\caption{EBSD map of delineations where the loading direction and\textit{ y} direction are aligned. (a) Phase map of \textalpha$_2$ and \textgamma. (b) Orientation map of \textalpha$_2$. (c) Orientation map of \textgamma.} \label{fig:EBSD}
\end{figure}

\begin{figure*}[t!]
\centering\includegraphics[width=1.8\columnwidth]{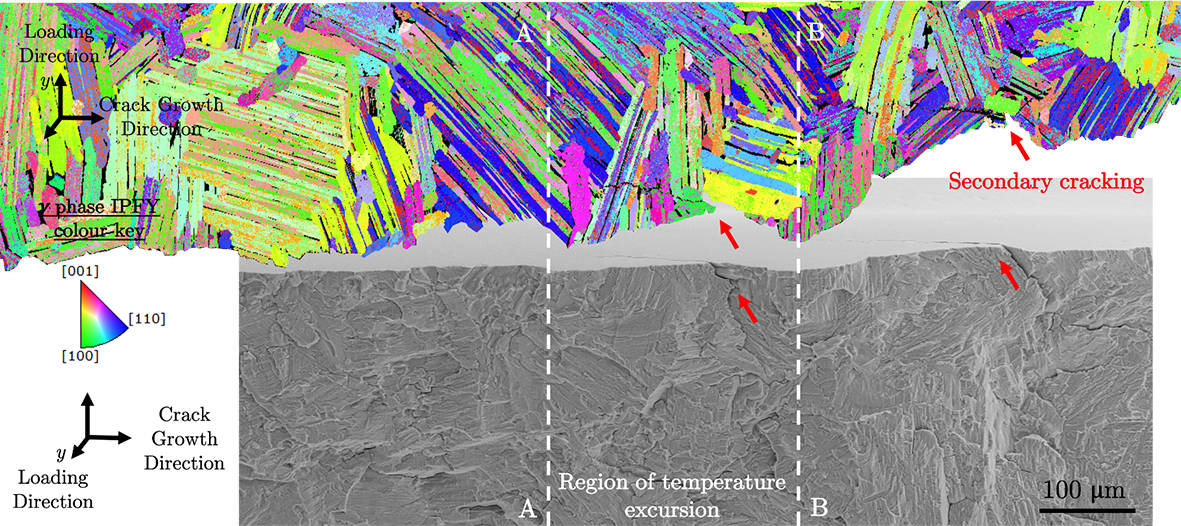}
\caption{EBSD map of the cross section perpendicular to the fracture surface through the region of the temperature excursion, with an SEM image below of the same region. The loading direction and \textit{y} direction are aligned up the page for the EBSD map.} \label{fig:2crackTE}
\end{figure*}

Figure \ref{fig:EBSD} is an EBSD map perpendicular to fracture surface delineations. Figure \ref{fig:EBSD}(a) shows the phase map, green is the \textgamma \hspace{0.1mm} phase, blue is the \textalpha$_2$ phase. Within a single colony \textalpha$_2$ has a single orientation due to it originating from the same parent \textalpha. It can therefore be deduced that the region depicted is a single colony due to Figure \ref{fig:EBSD}(b) showing the orientation of the \textalpha$_2$ being the same throughout. Figure \ref{fig:EBSD}(c) shows just the \textgamma \hspace{0.1mm} phase. This shows two orientations of the phase. However, because \textgamma--TiAl is only marginally tetragonal (2\% difference between \textit{a} and \textit{c}), Hough transform-based EBSD using only the major Hough peaks can find it difficult to uniquely assign an orientation. This is manifest as the assignment of green and pink IPF colours to what are clearly the same colony of $\gamma$ plates (which could be alternately twinned).

The fracture surface of test A was polished back perpendicular to the surface and an EBSD map (of the \textgamma \hspace{0.1mm} phase only) was generated of the region before, during and after the temperature excursion. This is shown in Figure \ref{fig:2crackTE} with an SEM image below of the corresponding perpendicular fracture surface. The SEM image shows no visible change in either morphology due to the increase in plastic zone size during the temperature excursion, or in contrast. The EBSD map shows no noticeable change through the temperature excursion with increased mechanical twinning or increased deformation. It can be seen that secondary cracking occurs in the reverse direction to the crack growth on the fracture surface. In this case, the secondary crack follows a grain boundary ahead, which may be related to stress relief in the plastic zone. In contrast, on inspection by light microscopy a marginally visible change in colouration could be observed due to, e.g. increased oxidation during the elevated temperature crack growth interval and the initial retardation of crack growth. \textcolor{red}{Examination of the other three specimens B-D both fractographically and by EBSD showed essentially similar behaviour.}

\subsection{Microstructural Initiation}
\begin{table}[b!]
\centering
\begin{small}\begin{tabular}{l|l|l}

\hline
\hspace{0mm}\textbf{Notch Size} & \hspace{5mm}\textbf{$\mathbf{\Delta K_{th}}$ } & \hspace{3mm}\textbf{$\mathbf{\Delta \sigma}$ } \\

\hspace{3mm}(mm) &  \hspace{2mm}(MPa$\sqrt{\mathrm{m}}$) &  \hspace{1mm}(MPa)\\
\hline
\hspace{4mm}0.03 &\hspace{6mm}n/a &\hspace{3mm}540\\
\hspace{4mm}0.06 &\hspace{6mm}n/a &\hspace{3mm}393\\
\hspace{5mm}0.1 &\hspace{6mm}4.0 &\hspace{3mm}331\\
\hspace{5mm}0.1 &\hspace{6mm}4.1&\hspace{3mm}340\\
\hspace{4mm}0.25 &\hspace{6mm}4.0&\hspace{3mm}209\\ \hline
\hspace{5mm}0.5 &\hspace{1mm}4.02$\pm$ 0.12&\hspace{1mm}147$\pm$ 5\\

\hline
\end{tabular}\end{small}
\caption{Measured effect of notch size on observed fatigue crack growth threshold at $450\celsius$; average over nine samples for 0.5mm notches.} \label{table:threshold}
\end{table}
Threshold determination was performed by gradually increasing $\Delta K$ rather than by pre-cracking under the $\Delta K$ decreasing method because the specimens were rather short and the crack growth rates in the materials rather fast. The choice of specimen geometry was a consequence of matching the casting conditions \textcolor{red}{and initial notch geometry} closely to those that might be expected to be used in aerofoils, i.e. limited sample section size \textcolor{red}{and foreign object damage}.  \textcolor{red}{There are also concerns with the $\Delta K$ decreasing method about loading history effects in \textgamma-TiAl alloys resulting in higher and non-conservative $\Delta K_\mathrm{th}$ values.} However, this then gave rise to a concern about what, if any, effect plastic deformation associated with machining of the notch might have. Therefore the initial notch size was decreased to determine if there was any variation in the observed $\Delta K_{th}$. Specimens with longer 0.1, 0.25 and 0.5 mm notches initiated from the notch through randomly oriented colonies, and show no effect from initial notch size on either trans-lamellar crack growth threshold or propagation, Table~\ref{table:threshold}. Nine values of trans-lamellar $\Delta K_{th}$ were measured for 0.5 mm, two values for 0.1 mm and one value for 0.25 mm\textcolor{red}{, all using the same machining disc and therefore possessing the same notch radius.}

\begin{figure}[b!]
\centering\includegraphics[width=1\linewidth]{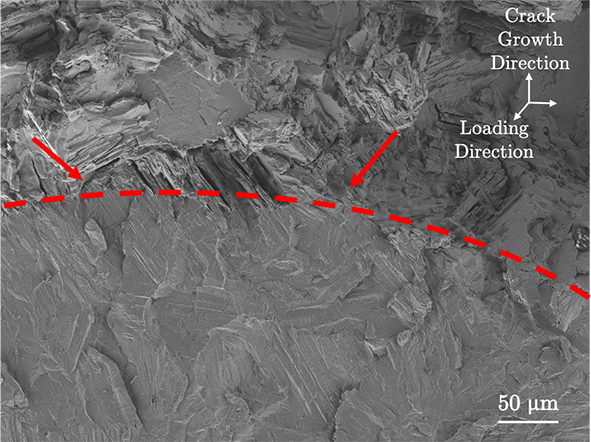}
\caption{Change in plastic zone size to larger than that of a typical colony causes the cracking to transition from wholly trans-lamellar to mixed inter-, intra- and trans- lamellar. This causes the roughness to abruptly increase. This transition point is shown by the red arrows.} \label{fig:PZ}
\end{figure}

\begin{figure}[b!]
\centering\includegraphics[width=1\linewidth]{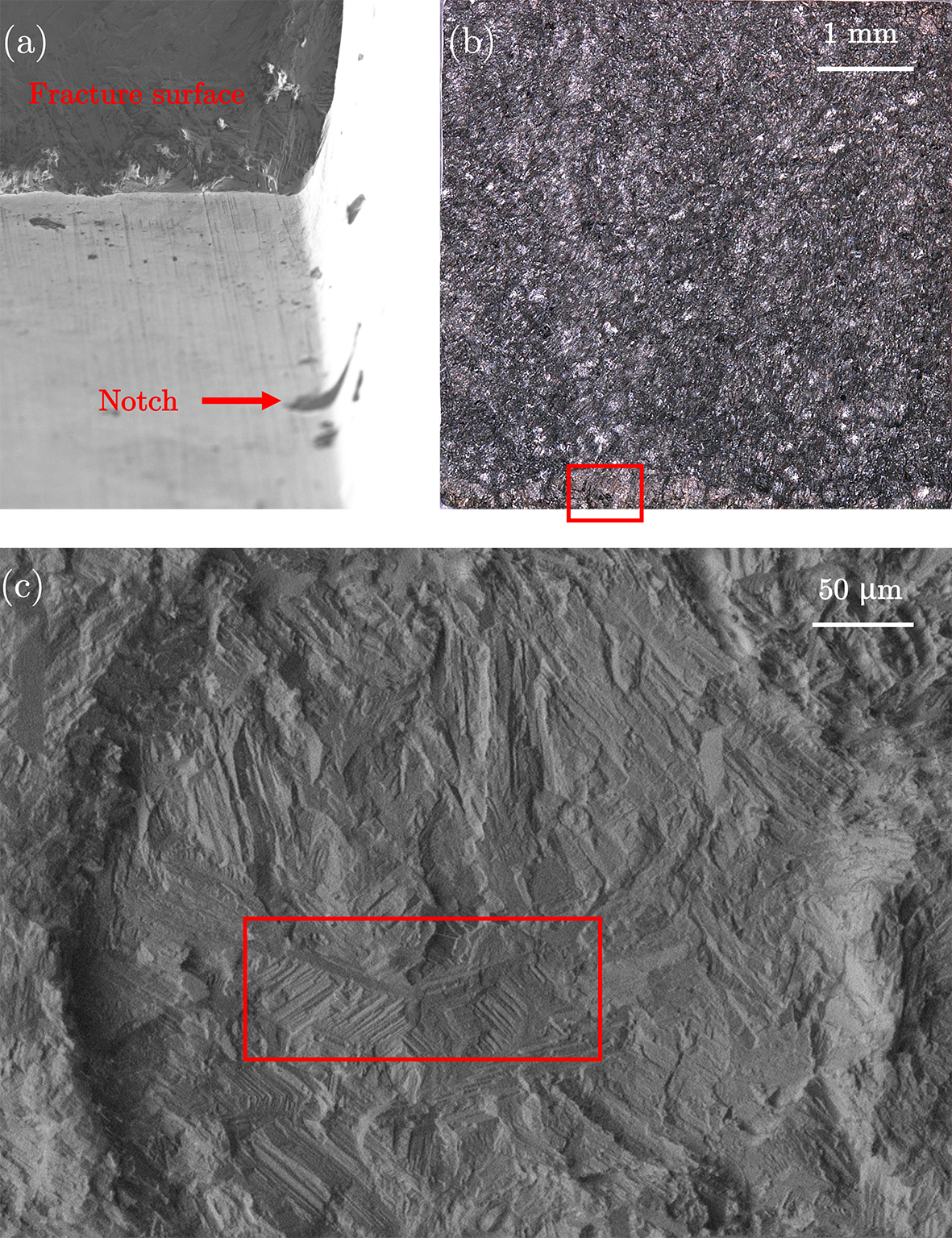}
\caption{The 0.03 mm notched specimen initiated away from the notch. (a)~The fracture surface can be seen at the top and the notch is indicated by the red arrow. (b) Fracture surface of the specimen with a small region of fatigue seen at the bottom of the micrograph, area of initiation marked by a red box. (c) Region of fatigue seen at the bottom of the fracture surface showing a near-circular, flatter region with trans-lamellar crack growth.} \label{fig:oo}
\end{figure}

Specimens with shorter 0.06 and 0.03~mm notches initiated naturally, cracking from microstructural features away from the notch. The maximum stress at which the 0.06 mm specimen initiated \textcolor{red}{(393 MPa, Table 2)} from the microstructure was close to the 0.1\% yield stress of 430 MPa (R=0.1), whilst in the 0.03 mm specimen \textcolor{red}{(with the FIB notch)} the uniaxial yield stress was exceeded. Therefore it is inferred that the natural microstructural features available to initiate failure are both consistent with uniaxial testing and imply that surface machining artefacts smaller than 0.06 mm might be regarded as benign.  \textcolor{red}{The lack of an effect of notch size on threshold and the existence of an intrinsic microstructural defect size is important for the assessment of vulnerability to, e.g. foreign object damage, which quite often only creates shallow indentations alongside potentially protective residual stress fields. However, the comparability of the damage fields associated with different types of flaw should be considered carefully in integrity assessment.} 

From the area of initiation, crack growth in fatigue was observed to be trans-lamellar, due to the plastic zone size being less than that of a colony, providing a flatter surface. With an increase in crack length, the stress intensity increases. Once this increased the plastic zone size to larger than the colony size, an increase in surface roughness could be observed, coinciding with a change in fracture mode to include inter- and intra-lamellar crack growth, Figure \ref{fig:PZ}.

\begin{figure}[t!]
\centering\includegraphics[width=1\linewidth]{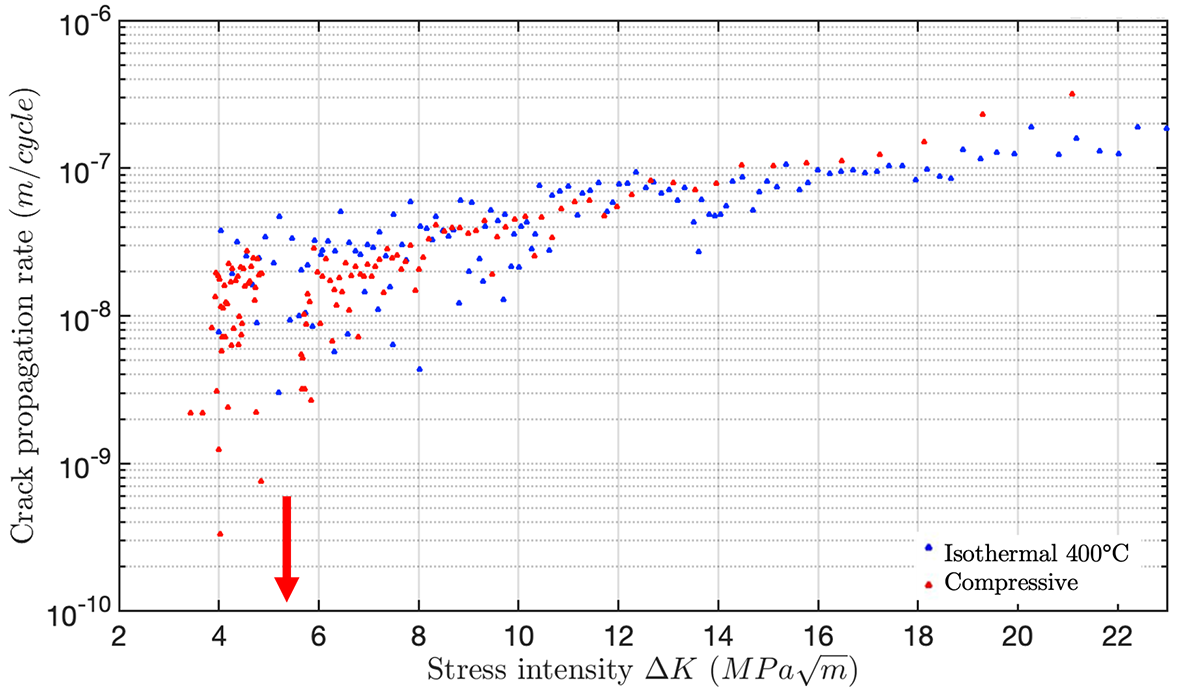}
\caption{Fatigue crack growth curve of the application of the compressive load of 6 kN at a $\Delta K$ of approximately 5 MPa$\sqrt{\mathrm{m}}$ indicated by the red arrow, compared to the isothermal response curve at 400\degree C.} \label{fig:comp}
\end{figure}

The specimen with the 0.03 mm FIB-machined notch initiated naturally from the central region of the gauge, but away from the notch which is indicated by the red arrows indicating the FIB-machined notch and fracture surface, Figure \ref{fig:oo}(a). A flatter region, indicating fatigue, could be observed at the bottom central side of the fracture surface, indicated by the red box in Figure \ref{fig:oo}(b). This region of fatigue, which exhibited a near-circular shape, can be seen in more detail in Figure \ref{fig:oo}(c). The roughness in this could be traced back to find an area of initiation\textcolor{red}{; with the benefit of experience, fractographic analysis is able to distinguish tensile overload from fatigue crack growth.} This initiation site is again observed to have a planar feature, with neighbouring intra-lamellar fracture, red box in Figure \ref{fig:oo}(c).

\subsection{Effect of a compressive event}
\begin{figure}[h!]
\centering\includegraphics[width=1\linewidth]{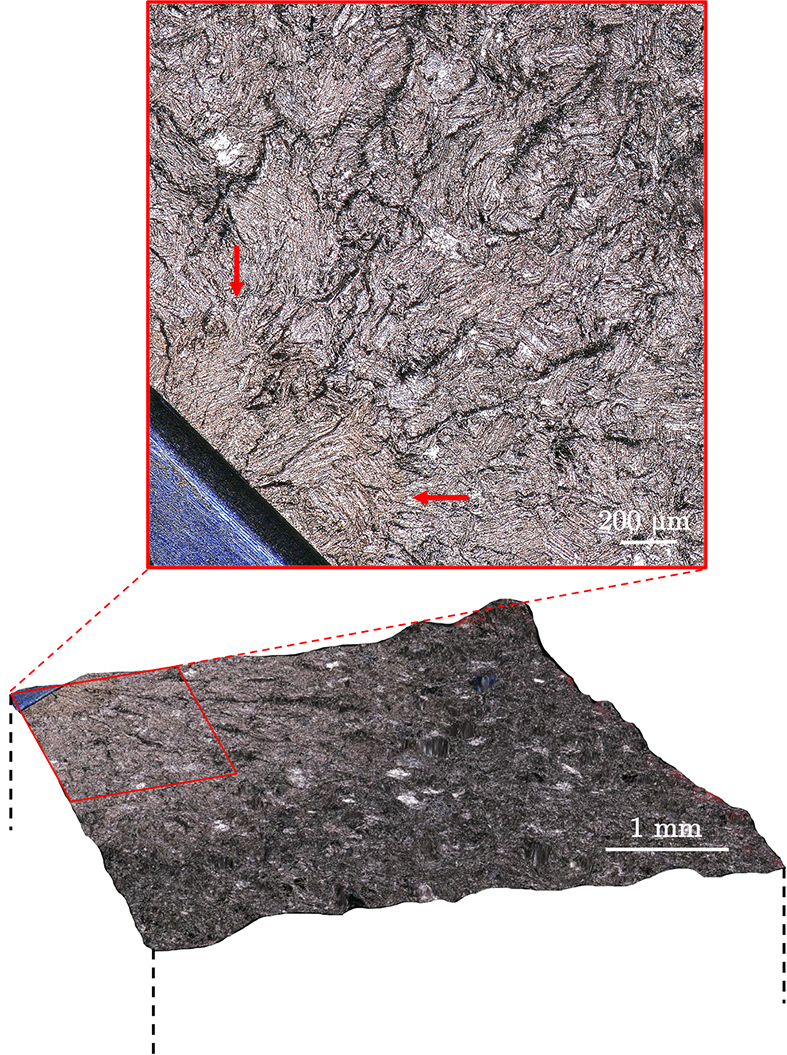}
\caption{Tilted fracture surface of compressive load application during crack growth with cropped inlay. There is a brown ring in the early crack growth marked by red arrows. This is oxidation due to being held at temperature for longer than rest of sample when growing crack out after the application of the compressive load.} \label{fig:compx}
\end{figure}

Application of a compressive load during crack growth causes crack retardation. Figure \ref{fig:comp} shows the initial crack growth following that of the isothermal response curve at 400\degree C. At a $\Delta K$ of approximately 5 MPa$\sqrt{\mathrm{m}}$ a compressive load of approximately 6 kN was applied (240 MPa). The crack growth rate could be seen to decrease, and then slowly grow out and return to the isothermal response curve. Figure \ref{fig:compx} shows the fracture surface of this specimen. This fracture surface was tilted to expose the surface morphology. A region of interest has been selected, cropped and rotated. There is a brown ring that can be seen in early crack growth marked by the red arrows. This is due to the region being held at temperature for longer as the crack retarded and took longer to grow out after the compressive event. This shows that there is no indication on the fracture surface of this compressive loading being applied, e.g. in fracture morphology or roughness, other than the region of increased oxidation. SEM fractographic examination and white light interferometry also provided no evidence to allow the compressive event to be discerned on the fracture surface.

% \begin{figure}[h]
% \centering\includegraphics[width=.7\linewidth]{compIMAGE.png}
% \caption{Fracture surface of the application of the compressive load during crack growth. There is a brown ring in the early crack growth marked by red arrows. This is due to oxidation because of being held at temperature for longer than rest of sample when trying to grow crack out after the application of the compressive load.} \label{fig:compI}
% \end{figure}

% \begin{figure}[h]
% \centering\includegraphics[width=1\linewidth]{compIMAGE2.png}
% \caption{Fracture surface at a tilt of the application of a compressive load during crack growth.} \label{fig:compx}
% \end{figure}

\section{Conclusions}
\textcolor{red}{At $400\celsius$ in air, cast and HIP'ed \textgamma-TiAl 4522XD has a trans-lamellar fatigue crack growth threshold of $\Delta K_\mathrm{th}=4.01\pm0.1\usk\mega\pascal\sqrt{\meter}$, whereas at $750\celsius$ trans-lamellar threshold was 7.1 MPa$\sqrt{\mathrm{m}}$. However, at $750\celsius$, higher fatgiue crack growth rates were observed.  Increases in temperature between these two conditions resulted in an initial burst of crack growth followed by crack retardation; decreases resulted in crack retardation alone.  Application of a compressive load during crack growth causes retardation in the crack growth.
For increases in temperature that left the crack below the trans-lamellar notched threshold of 7.1 MPa$\sqrt{\mathrm{m}}$, the existence of a sharp pre-crack resulted in inter-lamellar cracking with a lower threshold in air of 5.8 MPa$\sqrt{\mathrm{m}}$ at 750 \degree C.  In should be cautioned that threshold in \textgamma-TiAl alloys is found to be microstructure dependent and, for individual colonies, orientation dependent.
The temperature excursions could be clearly observed \emph{via} optical microscopy, through thickness changes in the nanometric oxide scale formed. However, changes in fractographic appearance or orientation gradient (in EBSD) could not be associated with the temperature excursions.  
%3) A step increase in temperature between the $\Delta K_\mathrm{th}$ values for the inter-lamellar and trans-lamellar threshold values for the increased temperature causes an initial burst in the crack growth rate, quickly followed by a retardation in the crack growth rate. The crack growth rate is found to have greater retardation the closer the value of $\Delta K$ at which the temperature is increased at is to the inter-lamellar fracture mode $\Delta K_\mathrm{th}$. 
%4) A step decrease in temperature above the $\Delta K_\mathrm{th}$ values shows crack growth rate retardation due to the larger plastic zone formed at the previously increased temperature. 
%5) The only method found in this work of detecting change in temperature is by optical microscopy, looking at the oxide scale colour. This oxide scale is not easily measurable, being less than a micron in thickness, in cross section. It not detectable by increased mechanical twinning in the region by looking at the EBSD maps obtained.
%6)
For specimens with a notch depth smaller than the size of a colony, $\sim$60 \micro m, microstructural initiation was observed. Specimens with notches larger than the colony size initiated from the notch. Notch size was not found to have an effect on the fatigue crack growth threshold. %The fatigue crack growth threshold for 0.5~mm starter notches was 4.02\pm0.12\hspace{0.1cm} $MPa\sqrt{\mathrm{m}}$ for nine specimens, 0.25~mm starter notch was 4.0~MPa$\sqrt{\mathrm{m}}$ for one specimen, and 0.1 mm starter notches was 4.0 and 4.1~MPa$\sqrt{\mathrm{m}}$ for two specimens. The fatigue crack growth threshold was therefore found to be 4.03\pm0.11 \hspace{0.1cm} $MPa\sqrt{\mathrm{m}}$ for the twelve above listed specimens. These specimens initiated in trans-lamellar fracture mode from the randomly oriented colonies local to the notch.
%8) Application of a compressive load during crack growth causes retardation in the crack growth.
Borides were seen to correspond with regions of initiation from microstructural features, but these secondary cracks implicated in the initiation or growth of the dominant crack.  Secondary cracking occurred across the fracture surfaces.}

\section{Acknowledgements}
This work was funded by EPSRC under a DTP CASE conversion, with specimens and additional funding provided by Rolls-Royce plc. The assistance of Dr Edward Saunders with the fractographic inspection and of Mr Jonathan Green with the macrophotography, both of Rolls-Royce plc., are gratefully acknowledged.

\bibliographystyle{model1-num-names}
\bibliography{references.bib}






% \end{thebibliography}

\end{document}